\documentclass[floatfix,onecolumn,showpacs]{article}

\usepackage{graphics,graphicx}
\usepackage{epsfig}
\usepackage{amsmath}
\usepackage{amstext}
\usepackage{amssymb}
\usepackage{dcolumn}

\newcommand{\beq}{\begin{eqnarray}}
\newcommand{\eeq}{\end{eqnarray}}

\begin{document}

\title{The nonadditive entropy $S_q$: A door open to the nonuniversality of the mathematical expression of the Clausius thermodynamic entropy in terms of the probabilities of the microscopic configurations\footnote{Invited review to appear in {\it Concepts and Recent Advances in Generalized Information
Measures and Statistics}, edited by A.M. Kowalski, R. Rossignoli and E.M.F. Curado (Bentham Science Publishers, 2011).}}

\author{Constantino Tsallis \\%}
% \affiliation{
Centro Brasileiro de Pesquisas Fisicas and \\ 
National Institute of Science and Technology for Complex Systems \\ 
Rua Xavier Sigaud 150, 22290-180 Rio de Janeiro-RJ, Brazil \\
and\\
Santa Fe Institute, 1399 Hyde Park Road, Santa Fe, NM 87501, USA}

\maketitle

\begin{abstract}
Clausius introduced, in the 1860´s, a thermodynamical quantity which he named {\it entropy} $S$. This thermodynamically crucial quantity was proposed to be {\it extensive}, i.e., in contemporary terms, $S(N) \propto N$ in the thermodynamic limit $N \to\infty$. A decade later, Boltzmann proposed a functional form for this quantity which connects $S$ with the occurrence probabilities of the microscopic configurations (referred to as {\it complexions} at that time) of the system. This functional is, if written in modern words referring to a system with $W$ possible discrete states, $S_{BG}=-k_B \sum_{i=1}^W p_i \ln p_i$, with $\sum_{i=1}^W p_i=1$, $k_B$ being nowadays called the Boltzmann constant ($BG$ stands for {\it Boltzmann-Gibbs}, to also acknowledge the fact that Gibbs provided a wider sense for $W$). The BG entropy is {\it additive}, meaning that, if A and B are two probabilistically independent systems, then $S_{BG}(A+B)=S_{BG}(A)+S_{BG}(B)$. These two words, {\it extensive} and {\it additive}, were practically treated by physicists, for over more than one century, as almost synonyms, and $S_{BG}$ was considered to be the unique form that $S$ could take. In other words, the functional $S_{BG}$ was considered to be universal. It has become increasingly clear today that it is {\it not} so, and that those two words are {\it not} synonyms, but happen to coincide whenever we are dealing with paradigmatic Hamiltonians involving {\it short-range} interactions between their elements, presenting no strong frustration and other ''pathologies". Consistently, it is today allowed to think that the entropic functional connecting $S$ with the microscopic world transparently appears to be {\it nonuniversal}, but is rather dictated by the nature of possible strong correlations between the elements of the system. These facts constitute the basis of a generalization of the BG entropy and statistical mechanics, introduced in 1988, and frequently referred to as nonadditive entropy $S_q$ and nonextensive statistical mechanics, respectively. We briefly review herein these points, and exhibit recent as well as typical applications of these concepts in natural, artificial, and social systems, as shown through theoretical, experimental, observational and computational predictions and verifications. 
\end{abstract}

%\pacs{???}

\maketitle

\section{Introduction}

In 1865, Rudolf Clausius introduced in thermodynamics a quantity $S$ that, in analogy with the word {\it energy}, he named {\it entropy}. The basic motivation was to introduce an integrating factor in the differential transfer of heat $\delta Q$, so that it would become an {\it exact differential}. The relation is now referred to as the Clausius equality for quasi-stationary process, namely $dS = \frac{\delta Q}{T}$, $T$ being the absolute temperature. Less than a decade later, Ludwig Boltzmann introduced a quantity which was defined in terms of the microscopic configurations of the system. This quantity gradually became identified as being a microscopic expression for the Clausius entropy $S$. We refer hereafter to this quantity as $S_{BG}$, where {\it BG} stands for {\it Boltzmann-Gibbs}, thus acknowledging the great contribution of Gibbs to the same concept. This functional form constitutes one of the cornerstones of the celebrated Boltzmann-Gibbs (BG) statistical mechanics. For a discrete set of probabilities $\{p_i\}$, it is given by
\begin{equation}
S_{BG}=-k \sum_{i=1}^W p_i \ln p_i = k \sum_{i=1}^W p_i \ln \frac{1}{p_i} \;\;\;\Bigl(\sum_{i=1}^W p_i=1 \Bigr)\,,
\label{entropydiscrete}
\end{equation}
where $W$ is the total number of possibilities, and $k$ a positive conventional constant chosen once for ever (the most frequent choices are $k=k_B$, or $k=1$).  If all probabilities are equal, we obtain the celebrated expression
\begin{equation}
S_{BG}=k \ln W\,.
\label{entropyquantum}
\end{equation}

If the variable which characterizes the microscopic state of the system is a $D$-dimensional continuous one ${\bf x}$, this expression is taken to be
\begin{equation}
S_{BG}=-k \int d{\bf x}\, p({\bf x}) \ln p({\bf x})= k\int d{\bf x}\, p({\bf x}) \ln \frac{1}{p({\bf x})} \;\;\;\Bigl(\int d{\bf x} \, p({\bf x})=1 \Bigr)\,,
\label{entropycontinuous}
\end{equation} 
and, if we are dealing with a quantum system, it is taken to be
\begin{equation}
S_{BG}=-k \,Tr\rho \ln \rho = k \,Tr\rho \ln \frac{1}{\rho} \;\;\;\Bigl(Tr \rho=1 \Bigr)\,,
\label{entropyquantum1}
\end{equation}
where $\rho$ is the density matrix. We shall indistinctively use here one or the other of these forms, depending on the particular point we are addressing. 

If our system is a dynamical one with an unique stationary state (a very frequent case), this state (referred to as {\it thermal equilibrium}, when we are dealing, in one way or another, with a macroscopically large amount of particles) is the one which, under appropriate contraints, maximizes $S_{BG}$. For example, for $D=1$, if we happen to know $\langle x \rangle \equiv \int dx\,x\, p(x)$, the maximizing distribution is given by
\begin{equation}
p(x)=\frac{e^{-\beta x}}{\int dx\, e^{-\beta x}} \;\;\;\ (\beta>0)\,,
\label{exponential}
\end{equation}
where $\beta$ is the corresponding Lagrange parameter.

If $\langle x \rangle =0$, and we happen to know $\langle x^2 \rangle$, the maximizing distribution is given by
\begin{equation}
p(x)=\frac{e^{-\beta x^2}}{\int dx\, e^{-\beta x^2}} \;\;\;(\beta>0)\,.
\label{gaussian}
\end{equation}
This Gaussian form is known to be consistent with the Maxwellian distribution of velocities for classical statistical mechanics, with the solution of the standard Fokker-Planck equation in the presence of a linear drift, and also with the Central Limit Theorem (CLT). The latter basically states that if we consider the sum $S_N=\sum_{i=1}^N X_i$ of $N$ {\it independent} (or nearly independent in some sense) random variables $\{X_i\}$, each of them having a {\it finite} variance, this sum converges for $N \to\infty$, after appropriate centering and rescaling, to a Gaussian. This  most important theorem can be proved in a variety of manners and under slightly different hypothesis. One of those standard proofs uses the Fourier transform, which we shall address later on.

It is well known that the BG entropy is the one to be used for a wide and important class of physical systems, basically those whose (nonlinear) dynamics is strongly chaotic ({\it positive} maximal Lyapunov exponent for classical systems), hence mixing, hence ergodic. What can be done with the others? In particular with those that might be weakly chaotic, with {\it vanishing} maximal Lyapunov exponent? It has been proposed in 1988 \cite{Tsallis1988} that the current statistical mechanical methods can be extended for an even wider class of physical systems just by generalizing $S_{BG}$. This generalization is reviewed in Section II. Before addressing it, let us make a crucial remark.

Let us consider a system whose total number of microscopic posibilities is given by $W^{(1)}$, hence, using Eq. (\ref{entropydiscrete}), we have that
\begin{equation}
S_{BG}^{(1)}=-k \sum_{i_1=1}^{W^{(1)}} p_{i_1}^{(1)} \ln p_{i_1}^{(1)}  \;\;\;\Bigl(\sum_{i_1=1}^{W^{(1)}} p_{i_1}^{(1)}=1 \Bigr)\,,
\label{entropydiscrete1}
\end{equation}
Let us assume next that we have $N$ such systems and that they are {\it probabilistically independent}, i.e., $p_{i_1,i_2,...,i_N}^{(N)}=p_{i_1}^{(1)}p_{i_2}^{(2)} ... p_{i_N}^{(N)} $, we straightforwardly prove that
\begin{equation}
S_{BG}(N)\equiv-k \sum_{i_1,i_2,...i_N}^{W(N)} p_{i_1,i_2,...,i_N}^{(N)} \ln p_{i_1,i_2,...,i_N}^{(N)}  \;\;\;\Bigl(\sum_{i_1,i_2,...,i_N}^{W(N)} p_{i_1,i_2,...,i_N}^{(N)}=1 \Bigr)\,,
\label{entropydiscreteN}
\end{equation}
is given by
\begin{equation}
S_{BG}(N)=S_{BG}^{(1)}+S_{BG}^{(2)}+...+S_{BG}^{(N)} \,,
\label{additivityN}
\end{equation}
where
\begin{equation}
W(N)=W^{(1)}W^{(2)}...W^{(N)} \,.
\end{equation}
It follows from Eq. (\ref{additivityN}) that $S_{BG}$ is {\it additive} \footnote{We are following the definition in \cite{Penrose1970} for entropic additivity, namely an entropy $S$ is {\it additive} if, for two probabilistically independent systems $A$ and $B$, $S(A+B)=S(A)+S(B)$. Otherwise it is {\it nonadditive}.}.

If the $N$ independent subsystems are equal we have that
\begin{equation}
W(N)=[W^{(1)}]^N \,,
\end{equation}
and
\begin{equation}
S_{BG}(N)=NS_{BG}(1) \propto N \,.
\end{equation}
Consequently the BG entropy of such a system is {\it extensive} since it satisfies the thermodynamical requirement that 
\begin{equation}
0<\lim_{N \to\infty}\frac{S_{BG}(N)}{N} < \infty\,, \;\;\; i.e., \;\; S(N) \propto N \;\;(N \to\infty)\,.
\label{extensivity}
\end{equation}
What happens if the $N$ subsystems (or elements) are not strictly independent but only nearly so? In other words, what happens if, for example, the system satisfies the following relation?
\begin{equation}
W(N) \sim \mu^N \;\;\;(N \to \infty; \;\mu>1) \,.
\label{BGsystem}
\end{equation}
This is the case, at infinite temperature, of the Ising model with first-neighboring ferromagnetic interactions in any Bravais lattice, of the Lennard-Jones gas, and of a plethora of other important systems. For all systems satisfying Eq. (\ref{BGsystem}), $S_{BG}$ is extensive since definition (\ref{extensivity}) is satisfied. 

But what happens if we have very {\it strong} generic correlations? More specifically, if the following relation is satisfied?
\begin{equation}
W(N) \sim N^\tau \;\;\;(N \to \infty; \;\tau>0) \,.
\label{qsystem}
\end{equation}
We immediately verify that $S_{BG}(N) = k\ln W(N) \propto \ln N$, hence it is {\it nonextensive} for any system satisfying (\ref{qsystem}). In other words, it violates classical thermodynamics. It is to avoid such disagreable situation that we are led to generalize the BG entropic functional.

\section{Generalizing Boltzmann-Gibbs entropy and statistical mechanics}
The generalization  we are focusing here is based on the following entropy \cite{Tsallis1988}:
\begin{equation}
S_{q}=k\frac{1-\sum_{i=1}^W p_i^q}{q-1}=-k \sum_{i=1}^W p_i^q \ln_q p_i = k \sum_{i=1}^W p_i \ln_q \frac{1}{p_i} \;\;\;\Bigl(\sum_{i=1}^W p_i=1 \Bigr)\,,
\label{qentropydiscrete}
\end{equation}

\begin{equation}
S_{q}=-k \int d{\bf x}\, [p({\bf x})]^q \ln_q p({\bf x})= k\int d{\bf x}\, p({\bf x}) \ln_q \frac{1}{p({\bf x})} \;\;\;\Bigl(\int d{\bf x} \, p({\bf x})=1 \Bigr)\,,
\label{qentropycontinuous}
\end{equation} 
and
\begin{equation}
S_{q}=-k \,Tr\rho^q \ln_q \rho = k \,Tr\rho \ln_q \frac{1}{\rho} \;\;\;\Bigl(Tr \rho=1 \Bigr)\,,
\label{qentropyquantum}
\end{equation}
where $q \in R$, and 
\begin{equation}
\ln_q z \equiv \frac{z^{1-q}-1}{1-q} \;\;\;(\ln_1 z=\ln z)\,.
\label{qlog}
\end{equation}
Eqs. (\ref{qentropydiscrete}), (\ref{qentropycontinuous}) and (\ref{qentropyquantum}) respectively recover Eqs. (\ref{entropydiscrete}), (\ref{entropycontinuous}) and (\ref{entropyquantum}) in the $q \to 1$ instance. This theory is currently referred to as {\it nonextensive statistical mechanics} \cite{Tsallis1988,CuradoTsallis1991,TsallisMendesPlastino1998,GellMannTsallis2004,BoonTsallis2005,Tsallis2009a,Tsallis2009b}. The entropy $S_q$ satisfies the following property: if $A$ and $B$ are two probabilistically {\it independent} systems, then
\begin{equation}
\label{sqadditive}
\frac{S_q(A+B)}{k}= \frac{S_q(A)}{k}+\frac{S_q(B)}{k}+(1-q)\frac{S_q(A)}{k}\frac{S_q(B)}{k}  \,.
\end{equation}
In other words, $S_q$ is {\it additive} \cite{Penrose1970} for $q=1$, and {\it nonadditive} for $q \ne 1$.  

Furthermore, if the probabilities are all equal, we have that
\begin{equation}
S_q=k \ln_q W \,.
\end{equation}

In remarkable contrast with $S_{BG}(N)$, $S_q (N) \propto N$ for $q=1-1/\tau$ for all systems satisfying (\ref{qsystem}). {\it For such complex systems, the additive entropy $S_{BG}$ is nonextensive, whereas the nonadditive entropy $S_q$ is extensive for a special (system-dependent) value of $q$, and therefore thermodynamically admissible!} Mathematical and physical examples can be seen in \cite{Tsallis2004,TsallisGellMannSato2005,CarusoTsallis2008,SaguiaSarandy2010}. We shall hereafter note $q_{ent}$ the value of $q$ such that $S_q(N) \propto N$. So, if the system satisfies relation (\ref{BGsystem}), we have $q_{ent}=1$, whereas if it satisfies (\ref{qsystem}) we have $q_{ent}=1-\frac{1}{\tau}$. It is convenient to emphasize at this point that particularly complex systems might exist (and probably do exist) for which no value $q$ exists such that $S_q$ is extensive.       

In the $D=1$ continuous case, if an appropriate first moment of $x$ is different from zero (see details in \cite{TsallisMendesPlastino1998}), the extremization of $S_q$ yields
\begin{equation}
p(x)=\frac{e_q^{-\beta x}}{\int dx\, e_q^{-\beta x}} \;\;\;(q<2;\, \beta>0)\,,
\label{qexponentialprobability}
\end{equation}
where $e_q^z$ is the inverse function of $\ln_q z$, more precisely
\begin{equation}
e_q^z \equiv [1+(1-q)z]_+^{\frac{1}{1-q}} \;\;\;(e_1^z=e^z) \,,
\label{qexponential}
\end{equation}
$[...]_+$ being equal to its argument when this is positive, and zero otherwise \footnote{To be more explicit, $e_q^z$ equals $e^z$ ($\forall z$) if $q=1$, vanishes for $z \le -1/(1-q)$ and equals $[1+(1-q)z]^{\frac{1}{1-q}}$ for $z>-1/(1-q)$ if $q<1$, and equals $[1+(1-q)z]^{\frac{1}{1-q}}$ for all $z<1/(q-1)$ (value at which it blows up to infinity) if $q>1$.}. Clearly Eq. (\ref{qexponentialprobability}) recovers Eq. (\ref{exponential}) for the $q=1$ particular case.

If the appropriate first moment vanishes and we happen to know the appropriate second moment, the extremization of $S_q$ yields, for the continuous case, the so-called $q$-Gaussian distribution
\begin{equation}
p(x)=\frac{e_q^{-\beta x^2}}{\int dx\, e_q^{-\beta x^2}} \;\;\;(q<3;\, \beta>0)\,,
\label{qgaussian}
\end{equation}

The $q$-Gaussian distributions \footnote{Since long known in plasma physics under the name {\it suprathermal} or {\it $\kappa$ distributions} \cite{RiosGalvao2010} if $q>1$, and equal to the Student's $t$-distributions \cite{SouzaTsallis1997} for special values of $q>1$.  They are also occasionally referred to as {\it generalized Lorentzians} \cite{Treumann1998}.} have a finite support for $q<1$, and an infinite one for $q \ge 1$; for $q>1$ they asymptotically decay as power-laws (more precisely like $x^{-2/(q-1)}$); they have a finite variance for $q<5/3$, and an infinite one for $q \ge 5/3$; they are normalizable only for $q<3$. 

Around 2000 \cite{BolognaTsallisGrigolini2000}, $q$-Gaussians have been conjectured (see details in \cite{Tsallis2005}) to be attractors in the CLT sense whenever the $N$ random variables that are being summed are strongly correlated in a specific manner. The conjecture was recently proved in the presence of $q$-independent variables \cite{UmarovTsallisSteinberg2008,UmarovTsallisGellMannSteinberg2010,NelsonUmarov2010,TsallisQueiros2007,QueirosTsallis2007}.  The proof presented in \cite{UmarovTsallisSteinberg2008} is based on a $q$-generalization of the Fourier transform, denoted as $q$-Fourier transform, and the theorem is currently referred to as the $q$-CLT. The validity of this proof has been recently challenged by Hilhorst \cite{Hilhorst2010}. His criticism is constructed on the inexistence of inverse $q$-Fourier transform for $q>1$, which he illustrates with counterexamples. The inverse, as used in the \cite{UmarovTsallisSteinberg2008} paper, indeed does not exist in general, which essentially makes the proof in \cite{UmarovTsallisSteinberg2008} a proof of existence, but not of uniqueness. The $q$-generalization of the inverse Fourier transform appears then to be a quite subtle mathematical problem. It has nevertheless been solved recently \cite{JaureguiTsallis2011}, and further considerations are coming \cite{JaureguiTsallisCurado2011} related to $q$-moments \cite{TsallisPlastinoAlvarezEstrada2009}. Work is in progress attempting to transform the existence proof in \cite{UmarovTsallisSteinberg2008} into a uniqueness one. In the meanwhile, several other forms \cite{VignatPlastino2007,HahnJiangUmarov2010} of closely related $q$-generalized CLT's have already been published which do {\it not} use the inverse $q$-Fourier transform.

Probabilistic models have been formulated \cite{RodriguezSchwammleTsallis2008,HanelThurnerTsallis2009} which, in the $N \to\infty$ limit, yield $q$-Gaussians. These models are scale-invariant, which might suggest that $q$-independence implies scale-invariance, but this is an open problem at the present time. However, definitively, scale-invariance does {\it not} imply $q$-independence. Indeed, (strictly or asymptotically) scale-invariant probabilistic models are known \cite{MoyanoTsallisGellMann2006,ThistletonMarshNelsonTsallis2009} which do {\it not} yield $q$-Gaussians, but other limiting distributions instead \cite{HilhorstSchehr2007}. In addition to all these, $q$-Gaussians are exact solutions of {\it nonlinear homogeneous} \cite{PlastinoPlastino1995,TsallisBukman1996,Borland1998,FuentesCaceres2008} as well as of {\it linear inhomogenous} Fokker-Planck equations \cite{AnteneodoTsallis2003}. In fact, these various cases have been recently unified as particular cases of {\it nonlinear inhomogenneous} Fokker-Planck equations \cite{MarizTsallis2011}. 

In Section 3 we remind some basic properties, and in Section 4 we briefly review various predictions, verifications and applications of $q$-exponentials and $q$-Gaussians through analytical, experimental, observational and computational methods in natural, artificial and social systems. However, before presenting them, let us make a few epistemological remarks.

Physics differs from mathematics. Within the traditional methodology of theoretical physics, mathematical tools are quite frequently (and fruitfully!) used {\it before} their complete mathematical understanding is achieved on rigorous grounds. In Henri Poincar\'e's words: {\it Deviner avant de d\'emontrer! Ai-je besoin de rappeler que c'est ainsi que se sont faites toutes les d\'ecouvertes importantes?} \footnote{{\it Guessing before proving! Need I remind you that it is so that all important discoveries have been made?}, Henri Poincar\'e, {\it La valeur de la science}, in Anton Bovier, {\it Statistical Mechanics of Disordered Systems}, (2006) page 218.}. 

For example, up to now there is no theorem determining the precise necessary and sufficient conditions for a classical or quantum many-body Hamiltonian to be legitimately admissible in order to know with mathematical certainty that its thermostatistical equilibrium state is given by the celebrated BG weight. Nevertheless, physics has exhibited amazingly interesting results and progress by so doing in a relatively {\it cavalier} manner! 

As a second example, we are not aware of any mathematical proof (rigorous by definition!) that the expansions based on the entire set of Feynman diagrams do converge in general to the exact answer. The same can be said for the successive truncations within the BBGKY  hierarchical scheme \cite{Balescu1975}.        
In fact these two latter illustrations most probably have some deep relation with the so-called {\it moment problem} in theory of probabilities. This problem focuses on the necessary and sufficient conditions for a probability distribution to be completely and uniquely determined by the complete set of its momenta, even when they are all {\it finite} (see \cite{TsallisPlastinoAlvarezEstrada2009,RodriguezTsallis2010} for cases where not all momenta are finite). Hundreds (maybe thousands) of useful developments in theoretical physics assume (sometimes explicitly, but more often just implicitly) that the full set of momenta perfectly determines the distribution. This question is so subtle that, in mathematics, three different moment problems are normally discussed separately for various classes of support of the probability distribution, namely for the $[0,1]$ support ({\it Hausdorff} moment problem), the $[0,\infty)$ support ({\it Stieltjes} moment problem), and the $(-\infty,\infty)$ support ({\it Hamburger} moment problem). General necessary and sufficient conditions are still ellusive \cite{math1,math2,math3,math4}.

However, the frequent lack of well established mathematical rigorous has never been a serious obstacle to the advancement of theoretical and experimental physics. We may once again quote Galileo:  {\it Nelle prove naturali non si deve ricercare l'esattezza geometrica} \footnote{{\it In the natural proofs we should not search for geometrical exactness.}}.

\section{Some basic properties of $S_q$}

We remind below some of the most basic properties of the $S_q$ 
entropy, besides the nonadditivity property shown in Eq. (\ref{sqadditive}). 
These properties are valid for all values of 
$q$, otherwise explicitly stated. For a more detailed list of the properties the reader can look at reference \cite{Tsallis2009a}. 

\begin{itemize}
  \item {\it Continuous dependence on the probabilites}: The $S_q$ function depends 
  continuously on the microscopic probabilities. 
  \item {\it Nul event}: The addition of a new event with zero probability doesn't change 
  the $S_q$ value for $q>0$. For $q<0$, by definition, 
  the sum in Eq. (\ref{qentropydiscrete}) runs only for states whose probability is positive.  
  \item {\it Certainty}: If one of the microscopic states has probability equal to one, the $S_q$ entropy equals to zero. 
  \item {\it Non-negativity}: $S_q(\{p_i \}) \geq 0$ for any set of probabilities $\{p_i \}$. 
  \item {\it Invariance with respect to the permutation of states}: $S_q$ given by Eq. (\ref{qentropydiscrete}) is invariant with respect to the permutation of all the states or, given by Eq. (\ref{qentropyquantum}), is invariant with respect to unitary transformations. 
 
  \item {\it Concavity}: The entropy $S_q$ satisfies the property $S_q(\{ p_i^{\prime \prime} \}) \ge 
 \lambda S_q(\{ p_i^{\prime } \}) + (1-\lambda) S_q(\{ p_i\} )$ for $p_i^{\prime \prime} = 
 \lambda p_i^{\prime} + (1-\lambda) p_i $, $\forall i$ ($0<\lambda<1$), i.e., it is 
 concave with respect to the probabilities for $q>0$.  For $q<0$ it is a convex function with 
 respect to the probabilities.
  \item {\it Extremum}: $S_q$ has its maximum value at equiprobability for $q>0$. 
  A simple proof can be seen by noticing that the function $s(p) = (p - p^q)/(q-1)$ is a 
  concave function and 
  that $S_q$ can be written like $\sum_i^W s(p_i)$. As $s(p)$ is a concave function, it satisfies 
  $s(\frac{1}{W}  \sum_i^W p_i ) = s(\frac{1}{W}) \geq \frac{1}{W} \sum_i^W s(p_i)$ for any 
  configuration $ \{ p_i \}$. For equiprobability we have 
  $s(\frac{1}{W} \sum_i^W \frac{1}{W}) = \frac{1}{W} \sum_i^W s(1/W) = s(1/W) \geq \frac{1}{W} \sum_i^W s(p_i)$,  
   for any configuration  $ \{ p_i \}$. The last inequality implies 
  that $ S_q(\{\frac{1}{W} \}) =  \sum_i^W s(1/W) = W s(1/W) \geq  \sum_i^W s(p_i) = S(\{ p_i \})$ for any configuration $\{ p_i \}$ other than equiprobability, showing that equiprobability extremizes $S_q$  \cite{Curado1999}. The same reasoning can be used to show that, for $q<0$, $S_q$ has a minimum 
  at equiprobability as now the function $s(p) = (p - p^q)/(q-1)$ is a convex one. 
  
  \item {\it Shannon additivity}: The Shannon additivity is slightly modified.
 % , satisfying now the property: $a$
 For a partition of the $W$ microscopic configurations into two subsets $A$ 
 and $B$, containing respectively $W_A$ and $W_B$ configurations satisfying 
 $W_A + W_B = W$, with respective probabilities $\{ p_1, \cdots, p_{W_A}  \}$ 
 and $\{ p_{W_A+ 1}, \cdots, p_{W}  \}$, the well-known Shannon additivity turns out 
 to be generalized as $S_q(p_1, \cdots, p_W) = S_q(p_A, p_B) + 
 p_A^q \, S_q(p_1/p_A, \cdots, p_{W_A}/p_A) + p_B^q \, S_q(p_{W_A+1}/p_B, 
 \cdots, p_W/p_B)$, where, in the above expression, $p_A=\sum_{i=1}^{W_A} p_i $ and 
 $p_B=\sum_{j=W_A+1}^{W} p_j $. 
 
 \item {\it Lesche stability}: The entropy $S_q$ is Lesche-stable (see details in \cite{Tsallis2009a}). 
 
\end{itemize}

\section{Applications}  

Let us list here some of the many verifications and applications of the present theory (i.e., of the nonadditive entropy $S_q$, and of its associated nonextensive statistical mechanics) that are available in the literature (see \cite{Bibliography} for full bibliography). (i) The velocity distribution of (cells of) {\it Hydra viridissima} follows
a $q$-Gaussian probability distribution function (PDF) with $q=3/2$  \cite{UpadhyayaRieuGlazierSawada2001}; (ii) The velocity distribution of (cells of) {\it Dictyostelium discoideum} follows a $q$-Gaussian PDF with $q=5/3$  in the vegetative
state and with $q=2$ in the starved state \cite{Reynolds2010}; (iii) The velocity
distribution in defect turbulence \cite{DanielsBeckBodenschatz2004}; (iv) The velocity
distribution of cold atoms in a dissipative optical lattice \cite{DouglasBergaminiRenzoni2006}; (v)
The velocity distribution during silo drainage \cite{ArevaloGarcimartinMaza2007a,ArevaloGarcimartinMaza2007b}; (vi) The
velocity distribution in a driven-dissipative 2D dusty plasma, with
$q=1.08\pm0.01$ and $q=1.05\pm 0.01$ at temperatures of $30000 \,K$
and $61000\, K$ respectively \cite{LiuGoree2008}; (vii) The spatial (Monte Carlo)
distributions of a trapped $^{136}Ba^+$ ion cooled by various
classical buffer gases at $300\,K$ \cite{DeVoe2009}; (viii) The distributions of
price returns and stock volumes at stock exchange, as well as the volatility smile  \cite{Borland2002a,Borland2002b,OsorioBorlandTsallis2004,Queiros2005}; (ix) Biological evolution \cite{TamaritCannasTsallis1998}; (x) The distributions of returns in the Ehrenfest's dog-flea model \cite{BakarTirnakli2009,BakarTirnakli2010}; (xi) The distributions of returns  in the
coherent noise model \cite{CelikogluTirnakliQueiros2010}; (xii) The distributions of returns of the
avalanche sizes in the self-organized critical Olami-Feder-Christensen model, as well as in real earthquakes \cite{CarusoPluchinoLatoraVinciguerraRapisarda2007};
(xiii) The distributions of angles in the $HMF$ model \cite{MoyanoAnteneodo2006}; (xiv) Turbulence in electron plasma \cite{AnteneodoTsallis1997}; 
 (xv) The relaxation in various paradigmatic spin-glass
substances through neutron spin echo experiments \cite{PickupCywinskiPappasFaragoFouquet2009}; (xvi) Various
properties directly related with the time dependence of the width of
the ozone layer around the Earth \cite{FerriReynosoPlastino2010}; (xvii) 
Various properties for conservative and dissipative nonlinear dynamical systems  
\cite{TsallisPlastinoZheng1997,BaldovinRobledo2002a,BaldovinRobledo2002b,Robledo2006,AnanosTsallis2004,BaldovinRobledo2004,MayoralRobledo2005,BorgesTsallisAnanosOliveira2002,TirnakliBeckTsallis2007,TirnakliTsallisBeck2009,TsallisTirnakli2010,AnteneodoTsallis1998,PluchinoRapisardaTsallis2007,PluchinoRapisardaTsallis2008,LyraTsallis1998,MiritelloPluchinoRapisarda2009,AfsarTirnakli2010,LeoLeoTempesta2010}; (xviii) The degree distribution of (asymptotically) scale-free
networks \cite{SoaresTsallisMarizSilva2005,ThurnerTsallis2005,WhiteKejzarTsallisFarmerWhite2006,ThurnerKyriakopoulosTsallis2007}; (xix) Tissue radiation response \cite{Sotolongo-GrauRodriguez-PerezAntoranzSotolongo-Costa2010}; (xx) Overdamped motion of interacting particles \cite{AndradeSilvaMoreiraNobreCurado2010}; (xxi) Rotational population in molecular spectra in plasmas \cite{ReisAmorimDalPino2011}; (xxii) High energy physics \cite{KaniadakisLavagnoQuarati1996,BediagaCuradoMiranda2000,Beck2000,TsallisAnjosBorges2003,Beck2003,WilkWlodarczyk2009,BiroPurcselUrmossy2009,CMS1,CMS2,EnterriaEngelPierogOstapchenckoWerner2011,PHENIX,ShaoYiTangChenLiXu2010,Wibig2010,AlbericoLavagnoQuarati2000}; (xxiii) Astrophysics \cite{CarvalhoSilvaNascimentoMedeiros2008, EsquivelLazarian2010,MoretSennaZebendeVaveliuk2010}; (xxiv) Analysis of the magnetic field in the solar wind plasma using data from Voyager 1 and Voyager 2 \cite{BurlagaVinas2005,BurlagaVinasAcuna2006,BurlagaNess2009}; (xxv) Nonlinear generalizations of quantum and relativistic equations \cite{NobreRegoMonteiroTsallis2011}.
The systematic study of metastable or long-living states in long-range versions of magnetic models such as the Ising \cite{NobreTsallis1995} and Heisenberg \cite{CarideTsallisZanette1983} ones, or in hydrogen-like atoms \cite{LucenaSilvaTsallis1995,OliveiraCuradoNobreMonteiro2007a,OliveiraCuradoNobreMonteiro2007b}, might provide further applications.

\section{Final remarks}

The BG entropy and its associated {\it exponential} distribution for thermal equilibrium have been extended, during the last two decades, in the sense of thermodynamics and statistical mechanics, into other forms. Stationary states have been discussed which correpond to the {\it $q$-exponential} form \cite{Tsallis1988}, {\it logarithmic} form \cite{Curado1999,CuradoNobre2004}, {\it stretched-exponential} form \cite{AnteneodoPlastino1999}, as well as other, more general, forms \cite{HanelThurner2011,Tempesta2011}. 

Let us make a comment on the results obtained in \cite{Tempesta2011}. By using group entropies, general forms of entropies are found which correspond to various Dirichlet {\it zeta} functions. The simplest of these functions is the Riemann {\it zeta} function, which turns out to correspond to the entropy $S_q$. This constitutes a rather striking and deep connection of nonextensive statistical mechanics with the theory of numbers. As a peculiarity it is also comes out in \cite{Tempesta2011} that the $q \to 1$ limit can not be straightforwardly implemented, thus making that the entropy $S_{BG}$ remains outside the class with real $q \ne 1$ (a remarkable mathematical fact which also occurs in \cite{HanelThurner2011}!).  

While by all means interesting, the connection with the Riemann $\zeta$ function established in \cite{Tempesta2011} is not a full surprise. Indeed, in 1993, the possibility of some kind of such connection was raised (see Figure) in a very particular context, namely the $q$-partition function of a single quantum harmonic oscillator. Let us review this point.
If we choose as zero energy the energy of the fundamental state, the spectrum of a single harmonic oscillator is given by
\begin{equation}
 E_n= \hbar \omega n \;\;\;\;(n=0,1,2,...) \,,
\end{equation}
hence the canonical $q$-partition function is given by
\begin{equation}
Z_q(\beta)= \sum_{n=0} e_q^{-\beta\,\hbar \omega n} \,.
\end{equation}  
For $q >1$, this expression becomes
\begin{eqnarray}
Z_q(\beta)&=& \sum_{n=0}^\infty \frac{1}{[1+(q-1)\beta\,\hbar \omega n]^{\frac{1}{q-1}}} \nonumber  \\ 
&=&\frac{1}{\Bigl[ (q-1) \beta \hbar \omega  \Bigr]]^{\frac{1}{q-1}}}\sum_{n=0}^\infty \frac{1}{\Bigl[\frac{1}{(q-1)\beta\,\hbar \omega}+n\Bigr]^{\frac{1}{q-1}}}  \nonumber  \\
&=&\frac{1}{\Bigl[ (q-1) \beta \hbar \omega  \Bigr]]^{\frac{1}{q-1}}}  \zeta\Bigl(\frac{1}{q-1},\frac{1}{(q-1)\beta\,\hbar \omega}\Bigr) \,,
\end{eqnarray} 
where $\zeta(s,Q)$ is the Hurwitz zeta function. The Riemann zeta function is given by
\begin{eqnarray}
\zeta(s)=\zeta(s,1)=\sum_{n=1}^\infty \frac{1}{n^s}=\prod_{p\;prime}\frac{1}{1-p^{-s}}=\frac{1}{1-2^{-s}}.\frac{1}{1-3^{-s}}.\frac{1}{1-5^{-s}}.\frac{1}{1-7^{-s}}.\frac{1}{1-11^{-s}}... \,,
\end{eqnarray}
which establishes a connection with the prime numbers, thus with the theory of numbers. We notice also that here the $q\to 1$ limit is pathological, like in \cite{HanelThurner2011,Tempesta2011}. Clearly, the partition functions $Z_q$ of systems different from the harmonic oscillator are not necessarily going to exhibit any relationship with the Riemman zeta function. It is nevertheless a kind of amusing coincidence that this function already appeared before in connection with $q$-statistics.  

As a final comment let us state that it appears that BG entropy and statistics are sufficient but not necessary for thermodynamics. In other words, thermodynamics might well be more powerful than the role attributed to it by BG statistical mechanics. A question which then arises naturally is: What is the most general form of entropy which can be consistent with thermodynamics, more precisely with the zeroth, first, second, and third principles? What are the most general superstatistical forms \cite{BeckCohen2003,TsallisSouza2003,HanelThurnerGellMann2011} which would correspond to this general entropy and statistics? By following in depth along the lines of \cite{HanelThurner2011,Tempesta2011,HanelThurner2011b}, it might be possible to clarify this important --- though hard --- question.

\begin{figure}
\begin{center}
\includegraphics[width=10cm]{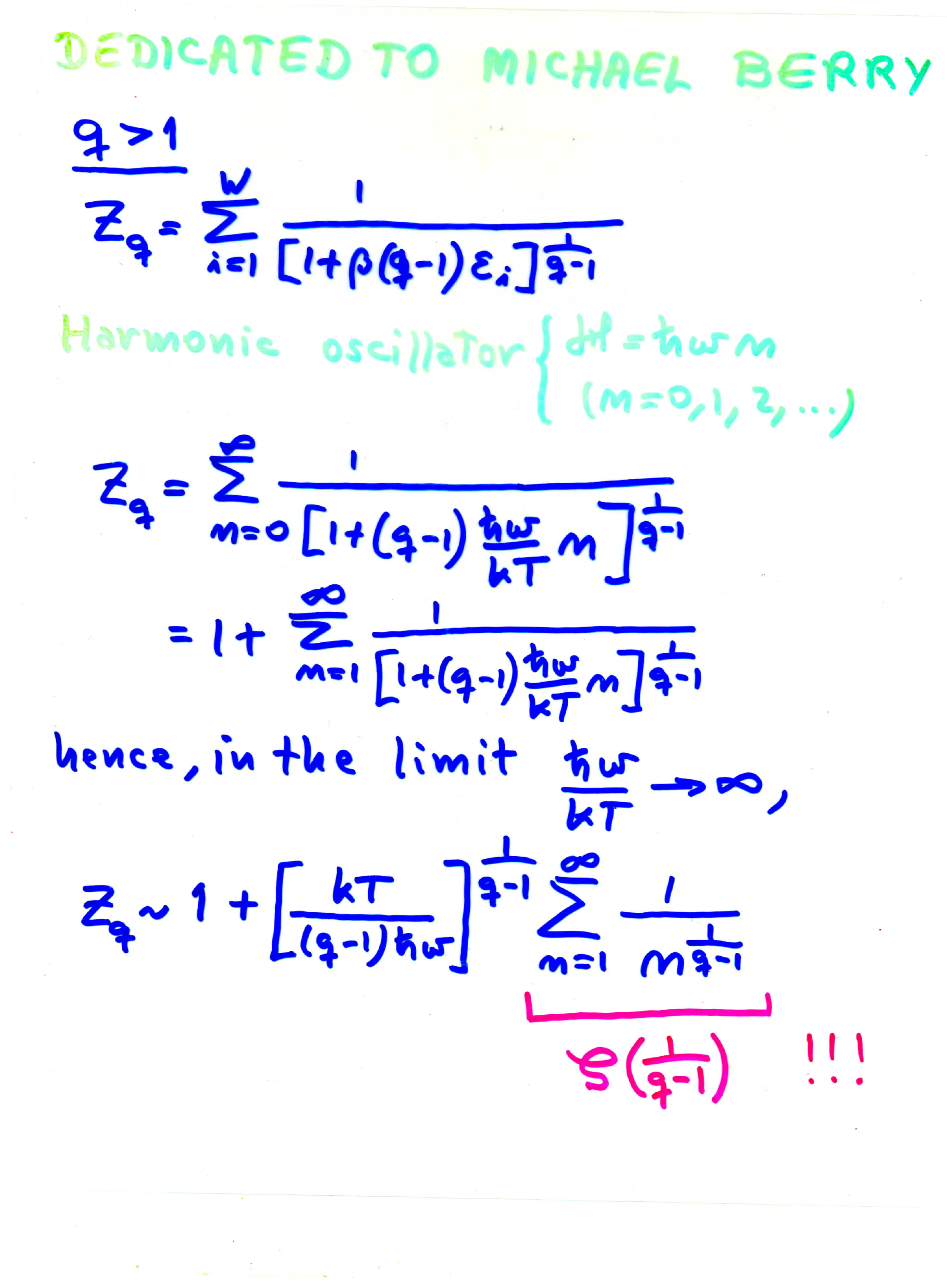}
\end{center}
\vspace{-2.0cm}
\caption{First occasion when it became apparent that there was a connection between $S_q$ (hence $q$-statistics) and the Riemann $\zeta$  function. I projected this slide during my lecture at the Workshop held, in 2-6 August 1993, in Mar del Plata, Argentina. The day before Michael Berry, in his own (wonderful) lecture, had emphasized that it would be very exciting if one could find fundamental connections of the prime numbers with physics. It is precisely such connection what the slide shows.}
\end{figure}

\section*{Acknowledgments}I acknowledge useful remarks by T. Bountis, E.M.F. Curado, R. Hanel, H.J. Hilhorst, M. Jauregui, F.D. Nobre, A. Rodriguez, G. Ruiz, P. Tempesta, S. Thurner and S. Umarov, as well as partial financial support by Faperj and CNPq (Brazilian agencies).

\end{document}